\documentclass[sigconf]{acmart} 

\usepackage{booktabs} 
\usepackage{multirow}
\usepackage{url}
\usepackage{algorithm}
\usepackage[noend]{algpseudocode}
\usepackage{amsmath}
\usepackage{color, colortbl} 
\usepackage{balance} 
\usepackage{pbox} 

\usepackage{graphicx}	
\usepackage{subcaption}

\setcopyright{rightsretained}

\acmDOI{10.475/123_4}

\acmISBN{123-4567-24-567/08/06}

\acmConference[Conference'18]{ACM Conference}{2018}{USA}
\acmYear{2018}
\copyrightyear{2018}

\acmArticle{4}
\acmPrice{15.00}

\definecolor{Gray}{gray}{0.9}
\definecolor{LightCyan}{rgb}{0.88,1,1}

\begin{document}
\title{How Much Are You Willing to Share? A \emph{``Poker-Styled''} Selective Privacy Preserving Framework for Recommender Systems}
%

\author{Manoj Reddy Dareddy}
\authornote{ 
Both authors contributed equally to this paper.
Author ordering is based on the alphabetical order of their last names, as suggested by the \href{http://www.ams.org/profession/leaders/culture/CultureStatement04.pdf}{American Mathematical Society (AMS)}.} 
\affiliation{
	\institution{Department of Computer Science\\
  	University of California, Los Angeles}
}
\email{mdareddy@cs.ucla.edu}

\author{Ariyam Das}
\authornotemark[1]
\affiliation{
	\institution{Department of Computer Science\\
  	University of California, Los Angeles}
}
\email{ariyam@cs.ucla.edu}

\author{Junghoo Cho}
\affiliation{
	\institution{Department of Computer Science\\
  	University of California, Los Angeles}
}
\email{cho@cs.ucla.edu}

\author{Carlo Zaniolo}
\affiliation{
	\institution{Department of Computer Science\\
  	University of California, Los Angeles}
}
\email{zaniolo@cs.ucla.edu}

\renewcommand{\shortauthors}{M. Dareddy et al.}
\renewcommand{\shorttitle}{A Selective Privacy Preserving Framework}
\renewcommand{\algorithmicrequire}{\textbf{Input:}}
\renewcommand{\algorithmicensure}{\textbf{Output:}}
\renewcommand{\algorithmiccomment}[1]{// #1}

%
%

\begin{abstract}
Most industrial recommender systems rely on the popular collaborative filtering (CF) technique 
for providing personalized recommendations to its users.
However, the very nature of CF is adversarial to the idea of user privacy, 
because users need to share their preferences with others in order to be grouped with like-minded people and receive accurate recommendations. 
Prior related work have proposed to preserve user privacy in a CF framework through different means like 
(i) random data obfuscation using differential privacy techniques, 
(ii) relying on decentralized trusted peer networks, or 
(iii) by adopting secured 
cryptographic strategies. 
While these approaches have been successful inasmuch as they 
concealed user preference information to some extent  
from a centralized recommender system,  
they have also, nevertheless, incurred significant trade-offs in terms of privacy, scalability, and accuracy. 
They are also 
vulnerable to privacy breaches by malicious actors.
In light of these observations, we propose a novel \textit{selective privacy preserving (SP2)} paradigm 
that allows users to custom define the scope and extent of their individual privacies, by marking their personal ratings as either \emph{public} (which can be shared) or \emph{private} (which are never shared and stored only on the user device).  
Our SP2 framework works in two steps: (i) First, it builds an initial recommendation model based on 
the sum of all \emph{public ratings} that have been shared by users and (ii) then, this public model is fine-tuned \emph{on each user's device} 
based on the user \emph{private} ratings, 
thus eventually learning a more accurate model. 
%
%
%

Furthermore, in this work, we introduce three different algorithms for implementing an end-to-end SP2 framework that can scale effectively from thousands to hundreds of millions of items.
Our user survey shows that 
an overwhelming fraction of users are likely to rate
much more items to improve the overall recommendations when they can control what ratings will be publicly shared with others. In addition, our experiments on two real-world dataset demonstrate
that SP2 can indeed deliver better recommendations than other state-of-the-art  methods, while preserving each individual user's self-defined privacy.

\end{abstract}

\maketitle

\section{Introduction}\label{intro}

Collaborative filtering (CF) based recommender systems are 
ubiquitously used across a wide spectrum of online applications ranging from e-commerce (e.g. Amazon) to recreation (e.g. Spotify, Netflix, Hulu, etc.) for delivering a personalized user experience \cite{mishra2016bottom}. 
CF techniques are broadly classified into two types -- (i) classic \emph{Nearest Neighbor} based algorithms \cite{MFNeighbor08} and  more recent \emph{matrix factorization techniques} \cite{MFNetflix},    
of which the latter has been more widely and predominantly 
adopted in industrial applications \cite{DasCIKM17} for building large-scale recommender models due to its superiority in terms of accuracy \cite{MFNetflix} and massive scalability \cite{FastSGDMFKDD15, MF16, Mahout, MLlib, Petuum, DiFacto}. 
Regardless of the underlying technique, the performance of a CF system is generally driven by the ``homophilous diffusion'' \cite{CannySIGIR02} process, where users must share some of their preferences in order to identify others with similar tastes and get good recommendations from them.
The performance of CF algorithms often deteriorates 
without such adequate information, as often observed in the classic \emph{cold start} \cite{ColdStart} 
problem.
%

This inherent need for a user to share his/her preferences sometimes leads to serious privacy concerns. To make things more complicated, 
privacy is not a static concept and may greatly vary across different users, items and places.  
%
For example, different users under changing geopolitical, social and 
religious influences may have varying degree of reservation about explicitly sharing their ratings on sensitive items that deal with subjects like politics, religion, sexual orientation, alcoholism, substance abuse, adultery, etc. \cite{ChowICDMW12}. 
%
Overall, these privacy concerns can prevent a user from explicitly rating many items, which reduces the overall performance of a CF algorithm, as compared to an ideal scenario, where everyone freely rates all the items they consume.

\subsection{Motivation}\label{motiv}
\begin{figure}
  \includegraphics[width=0.55\linewidth]{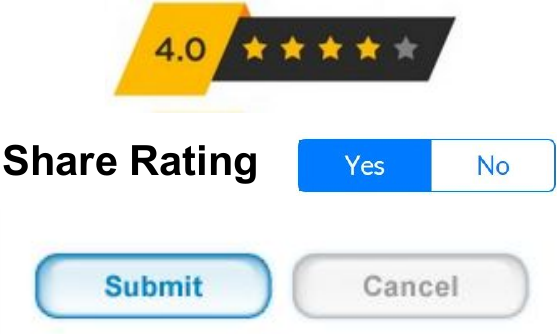}
  \caption{Working of a selective privacy preserving (SP2) framework from a user's perspective. 
  }
  \label{fig:sp2-gui}
\end{figure}

In this paper, we explore the idea of letting each user define his/her own privacy. In other words, here the user decides which 
ratings he/she can comfortably share 
\emph{publicly} with others, while his/her remaining ratings are considered as \emph{private}, which means that they are stored only on the user's device locally and are never shared with anyone including any peers or a centralized recommender system. Thus, this scheme enables each user to selectively define his/her own privacy. 
Figure \ref{fig:sp2-gui} shows an example of such an operational setup. 
%
%
In this paper, we attempt to build a CF framework that preserves each user's \emph{selective privacy} and investigates the following issues in enabling such a framework: 

\noindent $\bullet$ How can we build a \emph{selective privacy preserving} (\textbf{SP2}) CF model that assimilates information from two kinds of ratings -- all users' \emph{public} ratings and each user's on-device \emph{private} ratings? \vspace{2pt}

\noindent $\bullet$ How can we ensure that there is no loss of \emph{private} information in our SP2 framework? \vspace{2pt}

\noindent $\bullet$ Can the SP2 framework improve the performance of a CF algorithm? In other words, does the SP2 framework improve the overall recommendation quality at all by taking into account each user's private ratings? Or should the users simply hold back from rating sensitive materials if they have any privacy concern? \vspace{2pt}

%

\noindent $\bullet$ Can this SP2 CF model ensure scalability with respect to industrial-scale datasets? 

Interestingly, the selective privacy preserving framework proposed in this paper is somewhat analogous to 
the rules of a classic poker game\footnote{\href{https://en.wikipedia.org/wiki/Omaha_hold\_\%27em}{https://en.wikipedia.org/wiki/Omaha\_hold\_`em}} (\emph{Omaha hold `em}), where each player tries to form a best hand combining some of the community cards (which are publicly visible to everyone) and some of the hole cards (which are privately dealt to each player). 
\subsection{Contributions}\label{contri}
In the rest of this paper, we
address the questions listed in Section \ref{motiv} and make the following contributions: 

\noindent $\bullet$ We mathematically formulate the selective privacy preservation problem and present a formal framework to study it (Section \ref{architecture}). 
To the best of our knowledge, this is the first work 
under the umbrella of \emph{federated machine learning} \cite{OnDevice} 
that supports a \emph{private on-device} recommendation model 
for CF algorithms. \vspace{2pt}

\noindent $\bullet$ We propose three different strategies (Section \ref{approach}) for efficiently implementing an end-to-end SP2 framework, each of which is conducive to different situations. These underlying techniques overall ensure that a SP2 CF model incurs only a reasonable cost in terms of storage and communication overhead, even when dealing with massive industrial datasets or large machine learning models. 
\vspace{2pt}

\noindent $\bullet$ We present analytical results on two real datasets comparing different privacy preserving and data obfuscation techniques to show the effectiveness of our SP2 framework (Section \ref{experiments}).
We also empirically study 
what is a good information sharing strategy for any user in a SP2 framework and 
%
%
how much are the recommendations of a user affected, when he/she refrains from rating an item, instead of marking the latter as \emph{private}. \vspace{2pt}

\noindent $\bullet$ We 
present the results of 
a pilot study (Section \ref{survey}), which demonstrates 
that an overwhelming majority of participants are willing to adopt this technology in order to receive more relevant recommendations without sacrificing their individual privacies.  

\section{SP2 Architecture}\label{architecture}
Our proposed \emph{selective privacy preserving} (SP2) framework for CF algorithms is broadly based on the popular matrix factorization (MF) method, mainly due to its better performance, scalability and industrial applicability \cite{MFNetflix, MF16, Mahout, MLlib, DasCIKM17}. However, some of our discussions can also be extended to the traditional nearest neighbor based CF algorithms \cite{MFNeighbor08}. 
We next briefly review the MF technique in Section \ref{background}.
\subsection{Background}\label{background}
In the classic biased MF model \cite{MFNetflix}, we try to learn the latent user and item
factors (assumed to be in the same feature space of dimension $k$) from an incomplete ratings matrix \cite{MFNeighbor08}. More formally, here, 
the estimated rating for a user $u$ on item $i$, $\hat{r_{ui}}$ is given by equation (\ref{ratestimate}). The corresponding symbol  definitions are provided in Table \ref{tab:symbols}. 
We compute the user and item latent factors by minimizing the regularized squared error 
over all the known ratings, as shown in 
(\ref{l2loss}). 
This is done either using classic Alternating Least Squares method 
\cite{AlsRECSYS12,DasCIKM17,MLlib} 
which computes closed form solutions or via Stochastic Gradient Descent (SGD) \cite{MFNetflix},
which enjoys strong convergence guarantees 
\cite{NipsMFTheory,PMLRGDTheory} 
and many desirable properties for scalability 
\cite{FastSGDMFKDD15,AsyncSGD}. 
The variable update equations for SGD are given by equation (\ref{sgdupdates}). 
For simplicity, we assume from now on that the user and item factors contain the respective biases i.e. user factor for $u$ ($p_u'$) implies the column vector $[\, b_u \quad 1 \quad p_u^T\,]^T$ and item factor for $i$ ($q_i'$) 
refers to the column vector $[\, 1 \quad b_i \quad q_i^T\,]^T$. 
%
%
%
\begin{equation}\label{ratestimate}
\begin{aligned}
\hat{r_{ui}} = \mu + b_u + b_i + q_i^Tp_u = \mu + q_i'^Tp_u' 
\end{aligned}
\end{equation}
\begin{table}
  \caption{Definitions of symbols used in (\ref{ratestimate}) - (\ref{sgdupdates})}
  \label{tab:symbols}
  \begin{tabular}{llll}
    \toprule
    Symbol & Definition & Symbol & Definition\\
    \midrule
    $\mu$ & global mean of ratings &
    $\Omega$ & set of observed ratings\\
    $b_u$ & bias for user $u$ &
    $p_u$ & latent vector for user $u$\\
    $b_i$ & bias for item $i$ & 
    $q_i$ & latent vector for item $i$\\
    $\delta$ & Learning rate & 
    $\lambda$ & Regularization parameter \\    
    $r_{ui}$ & actual rating of $i$ by $u$ & 
    $\hat{r_{ui}}$ & prediction of $u$'s rating for $i$\\
    $e_{ui}$ & calculated as ($r_{ui}$-$\hat{r_{ui}}$) \\
  \bottomrule
\end{tabular}
\end{table}
\begin{equation}\label{l2loss}
\begin{aligned}
	\text{min} \sum\limits_{r_{ui}\in \Omega}{ (r_{ui}-\hat{r_{ui}})^2 + \lambda(b_i^2 + b_u^2 + \parallel q_i \parallel^2_2 + \parallel p_u \parallel^2_2 )}
\end{aligned}
\end{equation}
\begin{equation} \label{sgdupdates}
\begin{aligned}
    b_u &\leftarrow b_u + \delta(e_{ui} - \lambda b_u) \\
	b_i &\leftarrow b_i + \delta(e_{ui} - \lambda b_i) \\
	p_u &\leftarrow p_u + \delta(e_{ui}q_i - \lambda p_u) \\ 
	q_i &\leftarrow q_i + \delta(e_{ui}p_u - \lambda q_i)
\end{aligned}
\end{equation}
\subsection{Problem Formulation}\label{probform}
In a SP2 framework, each user $u$ has a set of \emph{public} ratings, denoted by $\Omega^u_{\text{public}}$ and a set of \emph{private} ratings, denoted by $\Omega^u_{\text{private}}$. 
However, since $\Omega^u_{\text{private}}$ is known only to $u$, the set of ratings observed here by the central recommender system is  $\bigcup\limits_{u} \Omega^u_{\text{public}}$. We denote the latter by the notation $\Omega_{\text{public}}'$. Now, our problem can be formulated as a \emph{multi-objective} optimization problem, where we attempt to minimize $n$ regularized L2 loss functions together for $n$ users, as shown below:\\

$\text{min}$ $(f_1, f_2, ..., f_n)$, where L2 loss ($f_v$) for user $v$ is given by,\\

$f_v: 
\big[ \sum\limits_{r_{vj}\in \Omega^v_{\text{private}}}{(r_{vj}-\hat{r_{vj}})^2} \big] 
+
\frac{1}{n}\sum\limits_{r_{ui}\in \Omega_{\text{public}}'}{ (r_{ui}-\hat{r_{ui}})^2}  \\ 
\verb|         | \qquad \qquad 
+ \frac{\lambda}{n}(b_i^2 + b_u^2 + \parallel q_i \parallel^2_2 + \parallel p_u \parallel^2_2 ) \\$

Note, traditionally multi-objective optimization problems are solved with classic techniques like \emph{linear scalarization} (also known as the weighted sum method \cite{linearScalarization}). In fact, 
if we assign equal weights to each user's L2 loss function, 
then linear scalarization \cite{linearScalarization} 
can reduce this problem into a single-objective mathematical optimization problem (constructed as the weighted sum of the individual objective functions), which is similar to the one discussed in Section \ref{background}. 
However, due to privacy considerations, all of the data (users' ratings) cannot be pooled together; this makes classic solutions to multi-objective optimizations problems inapplicable here. We next outline a privacy-aware model to solve this problem. 
\subsection{Model}\label{model}
We posit the following assumptions before summarizing our model. \\ 
\textbf{Assumption 1.}  The central recommender system is \emph{semi-adversarial} in nature i.e. it logs any information requested by a user and can later utilize it to guess what the user has rated privately. \\ 
\textbf{Assumption 2.} The central recommender system is \emph{not malicious} in nature i.e. it will not deliberately send incorrect information to a user to adversely impact his/her recommendations. It has an incentive to provide high quality recommendations to the users.\\
%
\noindent \textbf{Framework.} Based on the earlier discussions, 
we now outline the working of our SP2 framework:

\noindent $(1)$ The central recommender system first builds a \emph{public} model based on all the users' shared \emph{public} ratings using SGD. We obtain the \emph{public} user and item factors when the error converges after a certain number of epochs. 

\noindent $(2)$ Each user then downloads his/her corresponding \emph{public} user factor from the central recommender system. 

\noindent $(3)$ Additionally, all users' also download common \emph{auxiliary public model} data on their devices. This data is same for all users, and hence can be broadcasted by the central recommender system (for authentication in case the server cannot be trusted).  

\noindent $(3)$ Once the \emph{auxiliary public model} data 
and \emph{public} user factor is locally available on the device, local updates are performed on the \emph{public} user factor 
using \emph{auxiliary model} information and the \emph{private} ratings, which the user has saved on the device and has not shared with anyone.

\noindent $(4)$ The final \emph{private} user factor and the \emph{private} model 
are stored on the user's device and never shared or communicated. 
\begin{figure}
  \includegraphics[width=0.85\linewidth]{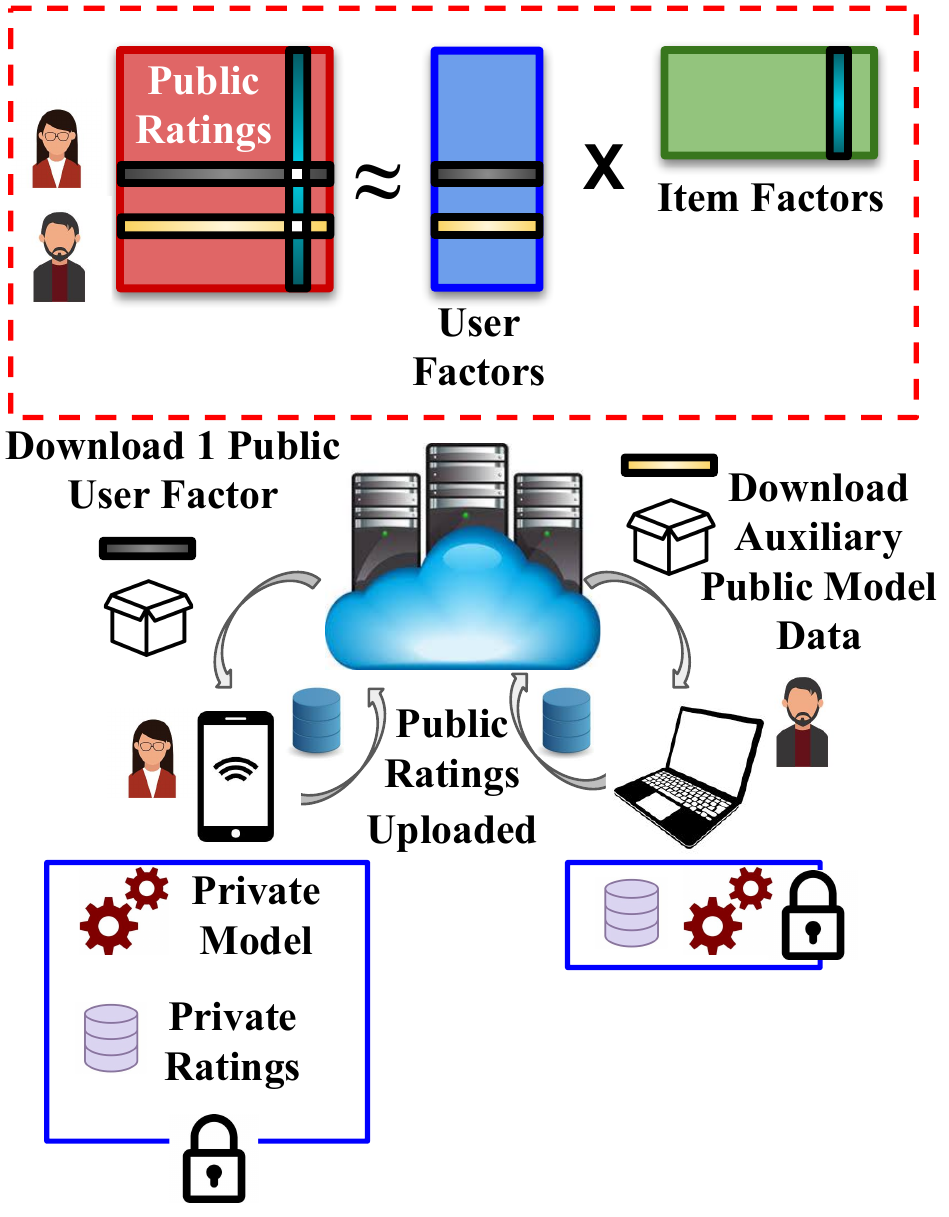}
  
  \caption{Architecture of a selective privacy preserving (SP2) framework.} \label{SP2Framework}
\end{figure}

Figure \ref{SP2Framework} presents the overall architecture. Interestingly, in our framework users never upload/communicate any \emph{private} rating, even in encrypted format, thus guaranteeing privacy preservation. 
This is notably different from the general federated machine learning philosophy \cite{OnDevice,SecuredAggProtocol}. We elaborate the need for this difference in Section \ref{related}.

\noindent \textbf{Auxiliary public model data.} It is also important to note that 
the central recommender system cannot share 
many parts of the \emph{public} model with all the users. 
In many systems like Netflix, users only consent to rate a video on the condition that the central recommender system displays only the average rating for the video, instead of the individual ratings from the users' community. 
In such cases, the \emph{auxiliary public model} data 
cannot contain ratings from $\Omega_{\text{public}}'$, as the latter scenario also constitutes a user-privacy breach, since the users may not be comfortable sharing their explicit ratings information with each other.
In the same vein, consider the example, 
where the \emph{auxiliary public model} data comprises of both \emph{public} user factors and item factors. This information alone is sufficient to identify the corresponding \emph{public} ratings for other users with reasonable confidence, thus again breaching user privacy. Furthermore, even anonymizing this information is not enough to prevent privacy leaks as demonstrated through de-anonymization attacks on Netflix dataset \cite{Deanonattacks}. 
Thus, the \emph{auxiliary public model} data needs to be designed carefully so that it 
not only facilitates in building a better \emph{private} model on the user's device, but also 
simultaneously 
safeguards the SP2 framework from privacy breaches. 
In light of this, observe that the \emph{auxiliary public model} data can comprise of \emph{public} item factors alone. Each {public} item factor is updated over a set of users based on their \emph{public} user factors and ratings. Thus, only the set of final \emph{public} item factors alone do not constitute a user-privacy breach.

\noindent \textbf{Private ratings distribution.} For analyzing the efficacy of our SP2 framework, it is also important to consider how users privately rate an item. We examine two different hypotheses for modeling this:

\noindent $\bullet$ \textbf{Hypothesis 1 (H1).} 
Users always decide independently which of his/her ratings are \emph{private}. Formally, for any two users $x$ and $y$, who have rated an item $i$ with ratings $r_{xi}$ and $r_{yi}$ respectively,  
$P(r_{xi}$ \text{is private} $|$ $r_{yi}$ \text{is private}$) = P(r_{xi}$ \text{is private}$)$.

\noindent $\bullet$ \textbf{Hypothesis 2 (H2).} 
Users do not decide independently which of his/her ratings are \emph{private}. In other words, ratings for some items are more likely 
to be marked as private. Formally, using the same mathematical notations as above, 
$P(r_{xi}$ \text{is private} $|$ $r_{yi}$ \text{is private}$) \neq P(r_{xi}$ \text{is private}$)$.

In Section \ref{experiments}, we further discuss how \emph{private} ratings are allocated in our experiments based on these two hypotheses.

\section{Implementation}\label{approach}
In this section, we present three approaches for implementing our SP2 framework. 
\subsection{Naive Approach}\label{naive}
In this approach, the auxiliary public model data contains the entire item factor matrix (i.e. all the latent item vectors and their biases). 
%
Each user's on-device \emph{private} model is then built following the steps shown in algorithm \ref{naivealgo}. The update equation used in this algorithm are similar to the ones used in the MF model in Section \ref{background}.
%
%
%
\begin{algorithm}
\caption{Naive method to build on-device \emph{private} model}
\label{naivealgo}
\begin{algorithmic}[1]
\Require 
$\delta \leftarrow$  learning rate ,
$\lambda \leftarrow$ reg. parameter ,
$epochs \leftarrow$ number of epochs,
$Q \leftarrow$ Aux. public model data containing all latent item vectors $(q_i)$, item biases $(b_i)$ and global ratings mean $(\mu)$,
$p_u \leftarrow$ \emph{public} user latent vector for $u$,
$b_u \leftarrow$ \emph{public} user bias for $u$,
$\Omega^u_{\text{private}} \leftarrow$ private ratings by $u$
\Ensure 
$p_u^* \leftarrow$ \emph{private} user latent vector for $u$,
$b_u^* \leftarrow$ \emph{private} user bias for $u$,
\Procedure{}{$\delta,\lambda,epochs, Q, p_u, b_u, \Omega^u_{\text{private}}$}
\State $p_u^* \leftarrow p_u, b_u^* \leftarrow b_u$
\For{$e=0;e<epochs;e++$}
	\ForAll{$ r_{ui}  \in \Omega^u_{\text{private}}$}
    	\State $\hat{r_{ui}} = \mu + b_u^* + b_i + q_i^Tp_u^*$
        \State $ e_{ui} = r_{ui}$-$\hat{r_{ui}}$
        \State $b_u^* \leftarrow b_u^* + \delta(e_{ui} - \lambda b_u^*) $
        \State $b_i \leftarrow b_i + \delta(e_{ui} - \lambda b_i) $
        \State $p_u^* \leftarrow p_u^* + \delta(e_{ui}q_i - \lambda p_u^*)$
        \State $q_i \leftarrow q_i + \delta(e_{ui}p_u^* - \lambda q_i)$
    \EndFor
    \State \textbf{end for}
\EndFor
\State \textbf{end for}
\EndProcedure
\end{algorithmic}
\end{algorithm}
\subsubsection{Top-$N$ recommendation}
Once the private model is built for user $u$, we can locally predict the rating for any item, as shown in equation (\ref{ratestimate}), using  $p_u^*, b_u^*$, since $q_i, b_i$ are known for all the items as part of the auxiliary public model data. These predictions can be ranked locally on the user device to provide the top-$N$ recommendations.   
%
%
\subsubsection{Privacy Consideration}
It is important to highlight some privacy considerations behind our naive approach:

\noindent $\bullet$ Even though a user only needs the corresponding item factors for each of the privately rated item to compute the on-device \emph{private} model, the user cannot simply fetch only the desired item factors from the central recommender system in order to ensure privacy.

\noindent $\bullet$ Consider an alternative scenario, where a user downloads only some additional irrelevant item factors to obfuscate the \emph{private} user information. This would require downloading significantly fewer number of item factors, as compared to downloading the entire item factor matrix.  However, this would make top-$N$ computation infeasible locally. Now, the user needs to send back $p_u^*, b_u^*$ to the server, which would allow the server to guess user's \emph{private} ratings. 
Similarly, sending a randomly perturbed \emph{private} user factor back to the server can obfuscate the \emph{private} information, but will degrade the quality of top-$N$ recommendations.

\noindent $\bullet$ Consider another alternative strategy, where the actual \emph{private} user factor is sent along with multiple ($k$) fake user factors, thereby obfuscating the private information and making it $k$-anonymous\cite{MicrosoftDiffPrivacy}. However, upload speeds are considerably lower than download speeds. In addition, the overall computation and communication costs can also increase by orders of magnitude, as the central servers need to compute multiple top-$N$ recommendation lists for every user and then send all of them back. 
%

It is important to note that the item factors matrix is downloaded only once during model building. In some situation, this does not involve unreasonable communication or storage overhead from the user end. For example, the total size (in MB) of all the item factors $(I)$ of dimension $k$ is given by $k \times |I| \times 8 / 2^{20}$, where each item factor is assumed to be an array of type \texttt{double}. Assuming $k=100$, the download sizes for all the item factors (in raw uncompressed format) for real datasets like MovieLens \cite{wsdm16} and Netflix \cite{wsdm16} are 4MB and 10MB respectively. However, for large industrial datasets (like Amazon \cite{mcauley2015image}) with close to 1 million items, the raw size of all item factors (of dimension 100) grows linearly to around 763MB.
%
\subsection{Clustering}
We propose this method to ensure scalability of the SP2 framework as the number of items become large. 
%
The intuition behind this approach is that the public auxiliary model data should consist of some approximate item factors $(Q')$, which is much smaller than the set of all item factors $(Q)$ i.e. $|Q'|<|Q|$. Now, each user $u$ for a private rating $r_{ui}$ should use the approximate item factor $\tilde{q_i'}$, instead of the actual item factor $q_i'$ to compute the \emph{private} model.  This approximation introduces an error in $e_{ui}$ calculation for each private rating $r_{ui}$ and is given by $p_u'^*(q_i' - \tilde{q_i'})^T$, 
where $p_u'^*$ is the \emph{private} user factor for $u$ and $\tilde{q_i'} \in Q'$.
Now, for each user $u$, we should minimize these approximation errors across all his/her \emph{private} ratings i.e. minimize 
$\sum_{i \in  \Omega^u_{\text{private}}}p_u'^*(q_i' - \tilde{q_i'})^T$, or
$p_u'^*\sum_{i \in  \Omega^u_{\text{private}}}(q_i' - \tilde{q_i'})^T$.
Since, the central recommender system does not know any $\Omega^u_{\text{private}}$ for any user, the former prepares the public auxiliary model data by minimizing the approximation errors across all item factors i.e. minimize  
$\sum_{i \in  Q}(q_i' - \tilde{q_i'})^T$. This minimization goal is similar to the objective function used in clustering \cite{KMeansClustering}. Thus, the central recommender system performs this approximation through clustering, particularly using $K$-means clustering with Euclidean distance \cite{kmeans++}. The individual cluster mean is treated as the approximate item factor for all the items in the cluster.
In summary, the public auxiliary model data for this method comprises of (1) $K$ cluster centroids obtained after applying the $K$-means algorithm on all the item factors, (2) cluster membership information, which identifies which cluster an item belongs to and (3) global ratings average. Using this public auxiliary model data, algorithm \ref{clusteringalgo} computes the on-device \emph{private} model for each user.  
\begin{algorithm}
\caption{Building on-device \emph{private} model via clustering}
\label{clusteringalgo}
\begin{algorithmic}[1]
\Require 
$\delta \leftarrow$  learning rate ,
$\lambda \leftarrow$ regularization parameter ,
$epochs \leftarrow$ number of epochs,
$Q' \leftarrow$ Aux. public model data containing all cluster centers having latent vectors $(c_i)$, biases $({b^c_i})$ and global ratings mean $(\mu)$,
$\rho \leftarrow$ Cluster membership function, where item $i$ is mapped to cluster $\rho(i)$,
$p_u \leftarrow$ \emph{public} user latent vector for $u$,
$b_u \leftarrow$ \emph{public} user bias for $u$,
$\Omega^u_{\text{private}} \leftarrow$ private ratings by $u$
\Ensure 
$p_u^* \leftarrow$ \emph{private} user latent vector for $u$,
$b_u^* \leftarrow$ \emph{private} user bias for $u$,
\Procedure{}{$\delta,\lambda,epochs, Q', p_u, b_u, \Omega^u_{\text{private}}$}
\State $p_u^* \leftarrow p_u, b_u^* \leftarrow b_u$
\For{\textbf{each} cluster c}
	\State $N_c = $ Calculate no. of items in $c$ from $\rho$ 
\EndFor
\State \textbf{end for}
\For{$e=0;e<epochs;e++$}
	\ForAll{$ r_{ui} $ in $ private\_ratings_u $}
    	\State $\hat{r_{ui}} = \mu + b_u + b^c_{\rho(i)} + c_{\rho(i)}^Tp_u^*$
        \State $ e_{ui} = r_{ui}$-$\hat{r_{ui}}$
        \State $b_u^* \leftarrow b_u^* + \delta(e_{ui} - \lambda b_u^*) $
        \State $b^c_{\rho(i)} \leftarrow b^c_{\rho(i)} + \delta(e_{ui} - \lambda b^c_{\rho(i)})/N_{\rho(i)} $ 
        \State $p_u^* \leftarrow p_u^* + \delta(e_{ui}c_{\rho(i)} - \lambda p_u^*)$
        \State $c_{\rho(i)} \leftarrow c_{\rho(i)} + \delta(e_{ui}p_u^* - \lambda c_{\rho(i)})/N_{\rho(i)}$
    \EndFor
    \State \textbf{end for}
\EndFor
\State \textbf{end for}
\EndProcedure
\end{algorithmic}
\end{algorithm}

Note, the cluster membership information for a set of $I$ items would require $4 \times |I|$ bytes, assuming each cluster id is an integer which takes 4 bytes. For $K$ clusters, this  membership information size can be further reduced drastically using $K$ bloom filters \cite{Bloom1970,BloomFilterWeb} where each bloom filter represents a cluster. 
%
\subsubsection{Top-$N$ recommendation}
In this method recall that all the item factors are not available locally on the user device. Therefore, we pursue a different strategy here: user $u$ requests the \emph{public} item factors for top-$N'$ recommended items $(N' > N)$ from the central recommender system. 
The latter computes this using $u's$ \emph{public} user factor $(p_u')$ and then sends the top $N'$ items and their corresponding \emph{public} item factors to $u$. $u$ can re-rank these $N'$ items based on his/her \emph{private} user factor $(p_u^*)$ and then select the top-$N$. Note, this top-$N'$ computation by the central servers is not a privacy threat, as it can be easily calculated without any information about user's private ratings. Also, recall our assumption 2 in Section \ref{model}, which ensures that incorrect top-$N'$ information will not be sent by the central servers. 
\subsection{Joint Optimization}
Our previous approach was based on hard assignment, where each item was assigned to only one cluster. However, soft clustering techniques like non-negative matrix factorization (NMF) \cite{MultUpdateProjGradDescent} considers each point as a weighted sum of different cluster centers. 
In this approach, we try to perform soft clustering on all the item factors simultaneously as the \emph{public} recommendation model is built. In other words, the central recommender system jointly learns the \emph{public} model and the soft cluster assignments.
For this, we revise the equations (\ref{ratestimate}) and (\ref{l2loss}) to 
(\ref{newratestimate}) and (\ref{newl2loss}), where 
$C$ denotes  
the cluster center matrix of dimension ${k \times z}$ ($z$ being the number of clusters), and $w_i$ is a column vector representing the different cluster weights (non-negative) for item $i$. 
This problem can be formulated as a constrained optimization problem and 
algorithm \ref{jointalgo} shows how the central recommender system performs this joint optimization. One key aspect in this algorithm is that the weights are updated (step 14) using projected gradient descent (PGD) \cite{ProjectedGradientDescent}, in order to ensure that all cluster weights are non-negative. This facilitates in finding the top-$R$ cluster assignments for any item by finding the highest $R$ corresponding weights. Finally, the auxiliary model data for this approach should consist of the following: (1) the cluster center matrix $C$, (2) item biases $b_i$, (3) top-$R$ cluster weights (in descending order) for each item $i$, the corresponding cluster ids and (4) the global ratings mean. Using $C$ and top-$R$ cluster weights for any item $i$, 
user $u$ can locally approximate the \emph{public} item factor for any item by its weighted sum of top-$R$ cluster centers 
i.e. $\sum\limits_{n \in \text{top $R$}}w_nC_n$ 
($C_n$ represents the $n^{th}$ cluster center). 
 With this approximation, $u$ can now use algorithm \ref{naivealgo} to compute the on-device \emph{private} model again. Note, when $R$ is small, we can save a significant communication cost by sending only top-$R$ weights as compared to the naive approach. 
\begin{equation}\label{newratestimate}
\begin{aligned}
\hat{r_{ui}} = \mu + b_u + b_i + w_i^TC^Tp_u 
\end{aligned}
\end{equation}
\begin{equation}\label{newl2loss}
\begin{aligned}
	\text{min} \sum\limits_{r_{ui}\in \Omega_{\text{public}}'}{ (r_{ui}-\hat{r_{ui}})^2 + \lambda(b_i^2 + b_u^2 + \parallel w_i\parallel^2_2 + \parallel c\parallel^2_2 + \parallel p_u\parallel^2_2 )}\\
    \text{s.t.} w_{ij} \geq 0.
\end{aligned}
\end{equation}
\begin{algorithm}
\caption{Joint optimization based matrix factorization}
\label{jointalgo}
\begin{algorithmic}[1]
\Require 
$\delta \leftarrow$  learning rate ,
$\lambda \leftarrow$ regularization parameter ,
$epochs \leftarrow$ number of epochs,
$\Omega_{\text{public}}' \leftarrow set of all \emph{public} ratings$
\Ensure 
$C, p_u, b_u, b_i, w_i$ for all users and items
\Procedure{}{$\delta,\lambda,epochs, \Omega_{\text{public}}'$}
\State $\mu = \text{Mean}(\Omega_{\text{public}}')$
\State Initialize $b_u, p_u, b_i, w_i, C$ with values from $N(0,0.01)$.
\For{$e=0;e<epochs;e++$}
	\ForAll{$ r_{ui} $ in $ private\_ratings_u $}
    	\State $\hat{r_{ui}} = \mu + b_u + b_i + w_i^TC^Tp_u$
        \State $ e_{ui} = r_{ui}$-$\hat{r_{ui}}$
        \State $b_u \leftarrow b_u + \delta(e_{ui} - \lambda b_u) $
        \State $b_i \leftarrow b_i + \delta(e_{ui} - \lambda b_i) $
        \State $p_u \leftarrow p_u + \delta(e_{ui}cw_i - \lambda p_u)$  
        \State $C \leftarrow C + \delta(e_{ui}w_ip_u^T - \lambda c)$
        \State $w_i \leftarrow w_i + \delta(e_{ui}c^Tp_u - \lambda w_i)$
        \For{\textbf{each} $w \in w_i$}
        	\State $w \leftarrow \text{Max}(w,0)$ //PGD
        \EndFor
        \State \textbf{end for}
    \EndFor
    \State \textbf{end for}
\EndFor
\State \textbf{end for}
\EndProcedure
\end{algorithmic}
\end{algorithm}
\subsubsection{Top-$N$ recommendation} Interestingly, with the auxiliary model data for this method, user $u$ can locally compute the approximation for each item factor, as mentioned above. As a consequence, $u$ is also able to locally compute the top-$N$ recommendations using these approximate item factors. 
\section{Experiments}\label{experiments}
We compared the performance of our SP2 framework with various baselines, as described next, under different settings on two real datasets, viz., MovieLens-100K \cite{harper2016movielens} data and a subset of Amazon Electronics \cite{mcauley2015image} data.
\subsection{SP2 vs. Different Baselines}
%
%
\noindent $\bullet$ \emph{Absolute Optimistic (Everything public)}: Here, we assume that every user optimistically shares everything publicly without any privacy concern i.e. a single MF model is built  on the entire training data itself. Theoretically, this should have the best performance, thus providing the overall upper bound.

\noindent $\bullet$ \emph{Absolute Pessimistic (Everything private)}: 
Here, we assume that every user is pessimistic and does not share anything publicly due to privacy concerns. Thus separate models are built for each user based \emph{only} on their individual ratings, which in practice, is as good as using the average rating for that user for all his/her predictions. 

\noindent $\bullet$ \emph{Only Public}: This mimics the standard CF scenario, where privacy preserving mechanisms are absent. Consequently, the users only rate the items, which they are comfortable with sharing; they refrain from explicitly rating sensitive items. 
We build a single MF model using \textbf{only} the \emph{public} ratings and ignore the \emph{private} ratings completely. 


\noindent $\bullet$ \emph{Distributed aggregation}: Shokri et al. \cite{ShokriRECSYS09} proposed peer-to-peer based data obfuscation policies, which obscured the user ratings information before uploading it to a central server that eventually built the final recommendation model. The three obfuscation policies mentioned are:
(1) \emph{Fixed Random (FR) Selection}: A fixed set of ratings are randomly selected from other peers for obfuscation.
(2) \emph{Similarity-based Random (SR) Selection}: A peer randomly sends a fraction of its ratings to the user for obfuscation depending on its similarity (Pearson, cosine, etc.) with the user.
(3) \emph{Similarity-based Minimum Rating (SM) Frequency Selection}: This approach is similar to the previous one, except that instead of randomly selecting the ratings, higher preference is given to the ratings of those items that have been rated the least number of times. 

\noindent $\bullet$ \emph{Fully decentralized recommendation}: Berkovsky et al. \cite{BerkovskyRECSYS07} proposed a fully decentralized peer-to-peer based architecture, where each user requests rating for an item by exposing a part of his/her ratings to a few trusted peers. The peers obfuscate their profiles by generating fake ratings and then compute their profile similarities with the user. Finally, the user computes the rating prediction for the item based on the ratings received from the peers and the similarities between them. 

\noindent $\bullet$ \emph{Differential Privacy}: McSherry et al. in \cite{MicrosoftDiffPrivacy} masks the ratings matrix sufficiently by adding random noise, drawn from a normal distribution, to generate a noisy global average rating for each movie. These global averages are then used to generate $\beta_{m}$ fictitious ratings to further obscure the ratings matrix. This method ensures that the final model obtained does not allow inference of the presence or absence of any user rating. 

For all MF models, the hyper-parameters were initialized with default values from the \texttt{Surprise} package\footnote{\url{http://surprise.readthedocs.io/en/stable/matrix\_factorization.html}}.
\subsection{Private Ratings Allocation}
We first provide the following two definitions, which are used later for the private ratings allocation:

\noindent $\bullet$ \emph{User privacy ratio} for a user $u$ is defined as the fraction of $u's$ total ratings which are marked \emph{private} by $u$.  

\noindent $\bullet$ \emph{Item privacy ratio} for an item $i$ is likewise defined as how many of the total users (which assigned $i$ a rating) have marked $i$ as \emph{private}. 

In order to examine the SP2 framework under two different hypotheses (stated in Section \ref{model}), we preprocess the datasets as discussed below:

\noindent $\bullet$ \textbf{H1.} We generate user privacy ratios in the interval $[0, 1]$ for all $n$ users from a beta distribution \cite{BetaDistr1} with parameters $\alpha, \beta$. For each user $u$ with user privacy ratio $\gamma_u$, $(1 - \gamma_u)$ fraction of $u$'s ratings are randomly selected and marked as \emph{public}, while the remainder of $u$'s ratings are considered \emph{private}.    

\noindent $\bullet$ \textbf{H2.} Here, we generate item privacy ratios for all $m$ items from a beta distribution. For each item $i$ with item privacy ratio $\gamma_i$, $(1 - \gamma_i)$ fraction of ratings assigned to $i$ are randomly selected and marked as \emph{public}, while the remainder of $i$'s ratings are considered \emph{private}. 

\begin{figure}
  \includegraphics[width=0.8\linewidth]{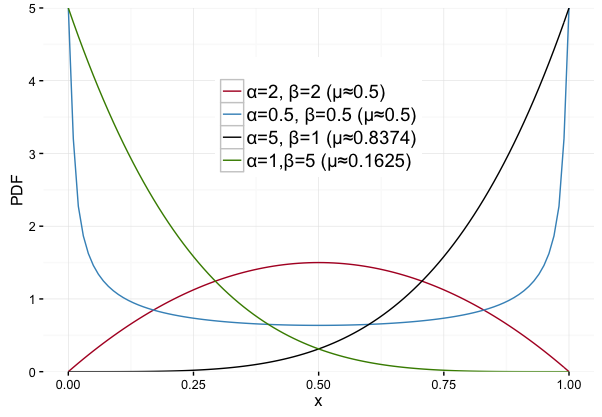}
  \caption{Probability density functions of four different beta distributions used in private ratings allocation.}
  \label{fig:BetaDists}
\end{figure}

For all our empirical analysis, we considered the following four beta distributions, as shown in Figure \ref{fig:BetaDists}. 

\noindent $1.$ \emph{Mostly Balanced $(\alpha = 2, \beta = 2)$}: Most user/item privacy ratios are likely to be close to the theoretical mean value 0.5.

\noindent $2.$ \emph{Mostly Extreme $(\alpha = 0.5, \beta = 0.5)$}: Most users/items have either very high or very low privacy ratios. The overall average of the privacy ratios will be close to 0.5.

\noindent $3.$ \emph{Mostly Pessimistic $(\alpha = 5, \beta = 1)$}: Most users/items have very high privacy ratios. 

\noindent $4.$ \emph{Mostly Optimistic $(\alpha = 1, \beta = 5)$}: Most users/items have very low privacy ratios. 
%
\subsection{Results}
We evaluate our SP2 framework using accuracy-based as well as ranking-based metric. The 5-fold average RMSE and NDCG@10 scores \cite{KDDNdcg} along with their corresponding standard deviations are reported in Table \ref{tab:results} for the MovieLens and Amazon Electronics datasets. 
%
%
%
\begin{table*}[ht]
\centering
\caption{Experimental results on MovieLens and Amazon Electronics datasets}
\label{tab:results}
\begin{minipage}{\textwidth} 
\begin{tabular}{lllcccc}
\toprule
\multirow{2}{*}{Category}             & \multirow{2}{*}{Method} & \multirow{2}{*}{Model Parameters} & \multicolumn{2}{c}{Movielens}                                               & \multicolumn{2}{c}{Amazon Electronics}                                       \\ 
                                      &                         &                             & RMSE                                 & NDCG@10                               & RMSE                                  & NDCG@10 \\ 
\midrule
\multirow{4}{*}{\pbox{20cm}{Peer-to-peer \\ Based}}         & Shokri et al. (FR)       &  \#Peers$ = 10$                           & 1.1624$\pm$0.00189                     & 0.4873$\pm$0.0055                    & 1.2216$\pm$0.01229                     & 0.7757$\pm$0.00841                      \\ 
                                      & Shokri et al. (SR)       & \#Peers$ = 10$                  & 1.1624$\pm$0.00562                     & 0.4891$\pm$0.00773                    & 1.2048$\pm$0.00889                      & 0.7774$\pm$0.00686                      \\ 
                                      & Shokri et al. (SM)       & \#Peers$ = 10$                 & 1.1447$\pm$0.00629                    & 0.4922$\pm$0.0094                    & 1.2028$\pm$0.00985                      & 0.7748$\pm$0.00887                      \\ 
                                      & Berkovsky et al.             &   \#Peers$ = 40$    & \multicolumn{1}{c}{1.132$\pm$0.00411} & \multicolumn{1}{c}{0.4876$\pm$0.00599} & \multicolumn{1}{c}{1.3405$\pm$0.00562} & \multicolumn{1}{c}{0.7619$\pm$0.00756} \\ 
   \midrule
   \midrule
Diff. Privacy                  &  McSherry et al.                        & $\beta_m = 15$     & 1.201$\pm$0.00675                      & 0.4795$\pm$0.00911                     & 1.1349$\pm$0.00664                      & 0.7719$\pm$0.00675                      \\ 
\midrule
\midrule
\multirow{2}{*}{\pbox{20cm}{Extreme \\ Baselines} 
} & {\color{red}Abs. Pessimistic}         & $k=100, \text{\#epochs}=20$                            & 0.9632$\pm$0.00489                     & 0.4132$\pm$0.00661                       & 0.9788$\pm$0.00368                      & 0.7379$\pm$0.00535                      \\ 
                                      & {\color{blue}Abs. Optimistic}          &       $k=100, \text{\#epochs}=20$      & 0.8923$\pm$0.00576                     & 0.5426$\pm$0.0072                         & 0.9538$\pm$0.00955                       & 0.788$\pm$0.00818                      \\ 
                                      \midrule \midrule
\multirow{8}{*}{
\pbox{20cm}{Classic \\ Collaborative \\ Filtering}
}                              & \multirow{4}{*}{Only Public (H1)}             &  
{\small$\alpha = 2, \beta = 2, \mu =0.48$}
& 0.9183$\pm$0.00725                     & 0.545$\pm$0.00726                     & 0.971$\pm$0.00516                      & 0.7892$\pm$0.00334                      \\ 
                              &              &  
{\small$\alpha = 0.5, \beta = 0.5, \mu =0.48$}
& 0.925$\pm$0.0075                    & 0.5468$\pm$0.00688                      & 0.9775$\pm$0.00455                      & 0.788$\pm$0.00204                      \\ 
                              &              &  
{\small$\alpha = 5, \beta = 1, \mu =0.82$}
& 0.9518$\pm$0.00822                     & 0.5363$\pm$0.00727                      & 0.9957$\pm$0.00763                      & 0.7738$\pm$0.00233                     \\ 
    &            	  &  
{\small$\alpha = 1, \beta = 5, \mu =0.17$} 
& 0.9033$\pm$0.00641                     & 0.5534$\pm$0.00179                      & 0.96$\pm$0.00411                      & 0.7955$\pm$0.00446                     \\ 
\cmidrule{2-7}
%
	& \multirow{4}{*}{Only Public (H2)}             &  
{\small$\alpha = 2, \beta = 2, \mu =0.48$}
& 0.9206$\pm$0.00328                     & 0.5391$\pm$0.00179                      & 0.9692$\pm$0.00895                      & 0.787$\pm$0.00463                      \\ 
                              &              &  
{\small$\alpha = 0.5, \beta = 0.5, \mu =0.48$}
& 0.9287$\pm$0.00228                     & 0.528$\pm$0.00596                     & 0.969$\pm$0.00931                      & 0.7853$\pm$0.00242                     \\ 
                              &              &  
{\small$\alpha = 5, \beta = 1, \mu =0.82$}
& 0.9522$\pm$0.00213                     & 0.517$\pm$0.00797                      & 0.9851$\pm$0.00808                      & 0.7718$\pm$0.00517                      \\ 
                              &              &  
{\small$\alpha = 1, \beta = 5, \mu =0.17$}
& 0.9063$\pm$0.00294                     & 0.5466$\pm$0.00212                     & 0.9581$\pm$0.00946                      & 0.7929$\pm$0.00456                      \\ 
\midrule \midrule
\multirow{24}{*}{\pbox{20cm}{\textbf{Selective} \\ \textbf{Privacy} \\ \textbf{Preserving} \\ \textbf{(SP2)}}}                              & \multirow{4}{*}{\textbf{Naive (H1)}}             &  
{\small$\alpha = 2, \beta = 2, \mu =0.48$}
& \textbf{0.9051$\pm$0.00654}                    & \textbf{0.5558$\pm$0.00511}                      & \textbf{0.9613$\pm$0.00534}                      & \textbf{0.7991$\pm$0.00322}                    \\ 
                              &              &  
{\small$\alpha = 0.5, \beta = 0.5, \mu =0.48$}
& \textbf{0.9072$\pm$0.00873 }                    & \textbf{0.5542$\pm$0.00727  }                    & \textbf{0.9641$\pm$0.00555 }                     & \textbf{0.7978$\pm$0.0012  }                    \\ 
                              &              &  
{\small$\alpha = 5, \beta = 1, \mu =0.82$}
& \textbf{0.9316$\pm$0.0088}                     & \textbf{0.5444$\pm$0.00666 }                     & \textbf{0.9808$\pm$0.00733 }                     & \textbf{0.7868$\pm$0.00297 }                     \\ 
                              &              &  
{\small$\alpha = 1, \beta = 5, \mu =0.17$}
& \textbf{0.8953$\pm$0.00688 }                    & \textbf{0.5594$\pm$0.00696 }                     & \textbf{0.9526$\pm$0.00525 }                     & \textbf{0.8048$\pm$0.00318 }                   \\ 
\cmidrule{2-7}
%
                             & \multirow{4}{*}{\textbf{Naive (H2)}}             &  
{\small$\alpha = 2, \beta = 2, \mu =0.48$}
& \textbf{0.907$\pm$0.00377}                     & \textbf{0.5514$\pm$0.00123 }                     & \textbf{0.9589$\pm$0.00921 }                     & \textbf{0.7977$\pm$0.00378 }                     \\ 
                              &              &  
{\small$\alpha = 0.5, \beta = 0.5, \mu =0.48$}
& \textbf{0.914$\pm$0.00302}                     & \textbf{0.5383$\pm$0.00491 }                     & \textbf{0.9603$\pm$0.00937 }                    & \textbf{0.793$\pm$0.00222  }                   \\ 
                              &              &  
{\small$\alpha = 5, \beta = 1, \mu =0.82$}
& \textbf{0.9316$\pm$0.00215}                     & \textbf{0.5274$\pm$0.00802 }                     & \textbf{0.9705$\pm$0.00795  }                    & \textbf{0.7824$\pm$0.00459  }                    \\ 
                              &              &  
{\small$\alpha = 1, \beta = 5, \mu =0.17$}
& \textbf{0.8946$\pm$0.0032}                     & \textbf{0.5532$\pm$0.00306 }                     & \textbf{0.9517$\pm$0.00949 }                    & \textbf{0.8034$\pm$0.00411 }                     \\ 
\cmidrule{2-7}
%
                             & \multirow{4}{*}{Clustering (H1)}             &  
{\small$\alpha = 2, \beta = 2, \mu =0.48$}
& 0.9165$\pm$0.00766                     & 0.5457$\pm$0.00695                      & 0.966$\pm$0.00565                      & 0.7893$\pm$0.00353\footnote{$P\text{value} < 0.02$}                      \\ 
                              &              &  
{\small$\alpha = 0.5, \beta = 0.5, \mu =0.48$}
& 0.9146$\pm$0.01183                     & 0.5494$\pm$0.00795                     & 0.9695$\pm$0.00774                      & 0.7876$\pm$0.00309                      \\ 
                              &              &  
{\small$\alpha = 5, \beta = 1, \mu =0.82$}
& 0.9387$\pm$0.00854                     & 0.5366$\pm$0.00681                      & 0.9847$\pm$0.00741                      & 0.7736$\pm$0.00241                      \\ 
                              &              &  
{\small$\alpha = 1, \beta = 5, \mu =0.17$}
& 0.9037$\pm$0.00634                    & 0.5538$\pm$0.00206                      & 0.958$\pm$0.0048                      & 0.7945$\pm$0.00373                    \\ 
\cmidrule{2-7}
%
                             & \multirow{4}{*}{Clustering (H2)}             &  
{\small$\alpha = 2, \beta = 2, \mu =0.48$}
& 0.9183$\pm$0.00395                    & 0.5401$\pm$0.00172                     & 0.9653$\pm$0.00924                     & 0.7871$\pm$0.00464\footnote{$P\text{value} < 0.1$, statistically insignificant}                      \\ 
                              &              &  
{\small$\alpha = 0.5, \beta = 0.5, \mu =0.48$}
& 0.9249$\pm$0.00206                     & 0.5287$\pm$0.00598                      & 0.9651$\pm$0.00938                      & 0.7852$\pm$0.00228                      \\ 
                              &              &  
{\small$\alpha = 5, \beta = 1, \mu =0.82$}
& 0.9405$\pm$0.00166                     & 0.5174$\pm$0.00718                      & 0.9757$\pm$0.00812                      & 0.7716$\pm$0.00528                     \\ 
                              &              &  
{\small$\alpha = 1, \beta = 5, \mu =0.17$}
& 0.9047$\pm$0.00319                     & 0.5473$\pm$0.00121                      & 0.9566$\pm$0.00971                      & 0.7926$\pm$0.00436                     \\ 
\cmidrule{2-7}
%
                             & \multirow{4}{*}{\textbf{Joint Opt. (H1)}}             &  
{\small$\alpha = 2, \beta = 2, \mu =0.48$}
& \textbf{0.9051$\pm$0.00654 }                    & \textbf{0.556$\pm$0.00502 }                     & \textbf{0.9612$\pm$0.00533  }                    & \textbf{0.7989$\pm$0.00315 }                     \\ 
                              &              &  
{\small$\alpha = 0.5, \beta = 0.5, \mu =0.48$}
& \textbf{0.9072$\pm$0.00873 }                    & \textbf{0.5537$\pm$0.00735 }                     & \textbf{0.964$\pm$0.00556  }                    & \textbf{0.7975$\pm$0.00098 }                    \\ 
                              &              &  
{\small$\alpha = 5, \beta = 1, \mu =0.82$}
& \textbf{0.9316$\pm$0.0088  }                   & \textbf{0.5447$\pm$0.00646 }                    & \textbf{0.9808$\pm$0.00734 }                     & \textbf{0.7869$\pm$0.00338 }                     \\ 
                              &              &  
{\small$\alpha = 1, \beta = 5, \mu =0.17$}
& \textbf{0.8953$\pm$0.00689}                    & \textbf{0.5592$\pm$0.00706 }                     & \textbf{0.9526$\pm$0.00524 }                    & \textbf{0.8045$\pm$0.00318 }                     \\ 
\cmidrule{2-7}
%
                             & \multirow{4}{*}{\textbf{Joint Opt. (H2)}}             &  
{\small$\alpha = 2, \beta = 2, \mu =0.48$}
& \textbf{0.907$\pm$0.00377}                     & \textbf{0.551$\pm$0.00095}                      & \textbf{0.9589$\pm$0.0092}                      & \textbf{0.7978$\pm$0.00371}                      \\ 
                              &              &  
{\small$\alpha = 0.5, \beta = 0.5, \mu =0.48$}
& \textbf{0.914$\pm$0.00302}                     & \textbf{0.5383$\pm$0.00507}                      & \textbf{0.9602$\pm$0.00939}                      & \textbf{0.793$\pm$0.0022}                      \\ 
                              &              &  
{\small$\alpha = 5, \beta = 1, \mu =0.82$}
& \textbf{0.9319$\pm$0.00241}                     & \textbf{0.5278$\pm$0.00851 }                     & \textbf{0.9705$\pm$0.00792 }                     & \textbf{0.782$\pm$0.00479 }                     \\ 
                              &              &  
{\small$\alpha = 1, \beta = 5, \mu =0.17$}
& \textbf{0.8947$\pm$0.0032 }                    & \textbf{0.5533$\pm$0.00289 }                     & \textbf{0.9517$\pm$0.0095 }                     & \textbf{0.8034$\pm$0.00423 }                     \\ 
\bottomrule
\end{tabular}
\end{minipage} 
\end{table*}
\begin{figure}
   \begin{subfigure}[b]{0.48\linewidth}
        \includegraphics[width=\linewidth]{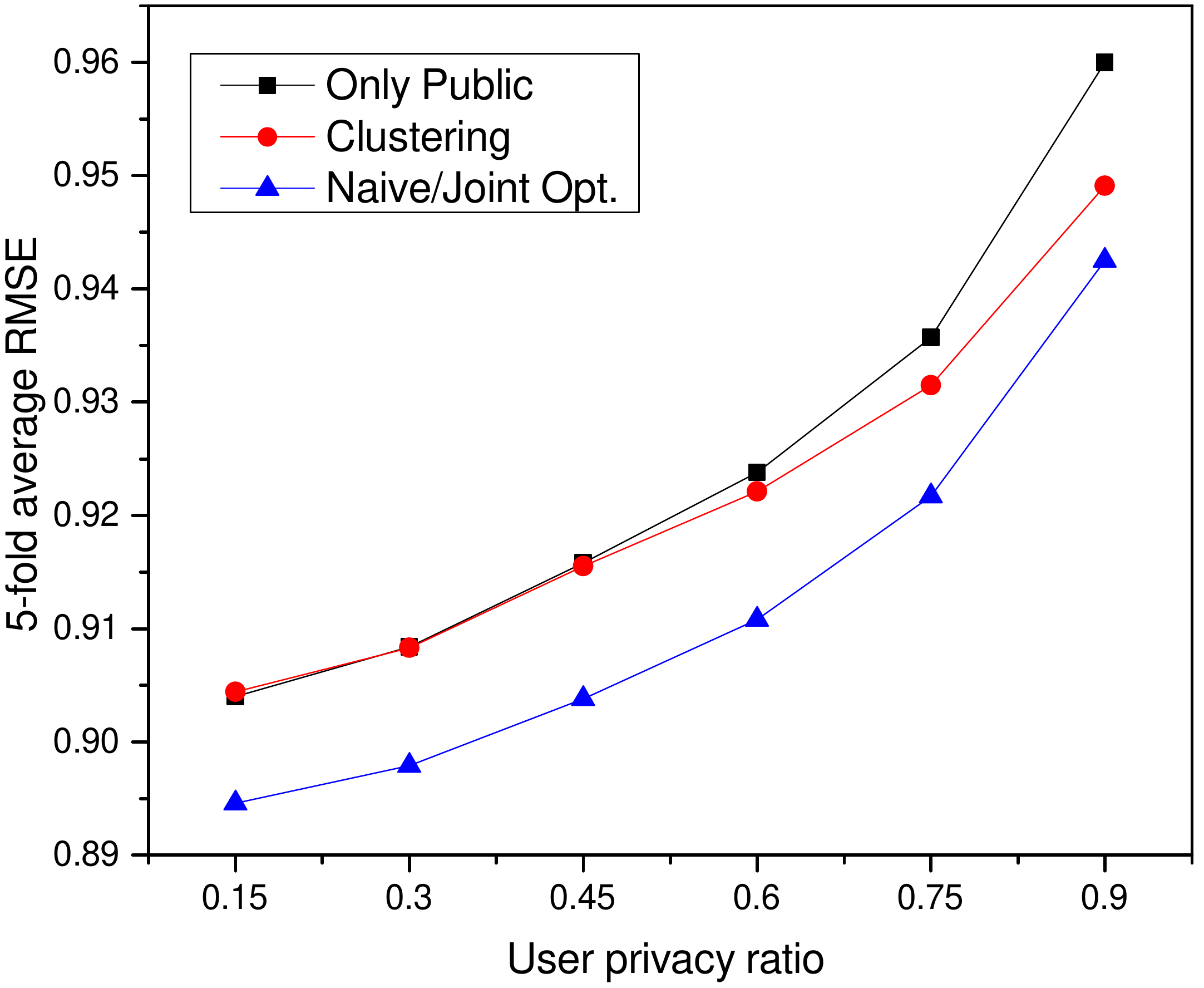}
        \caption{RMSE comparison}
        \label{fig:up_rmse}
   \end{subfigure}
    \begin{subfigure}[b]{0.48\linewidth}
        \includegraphics[width=\linewidth]{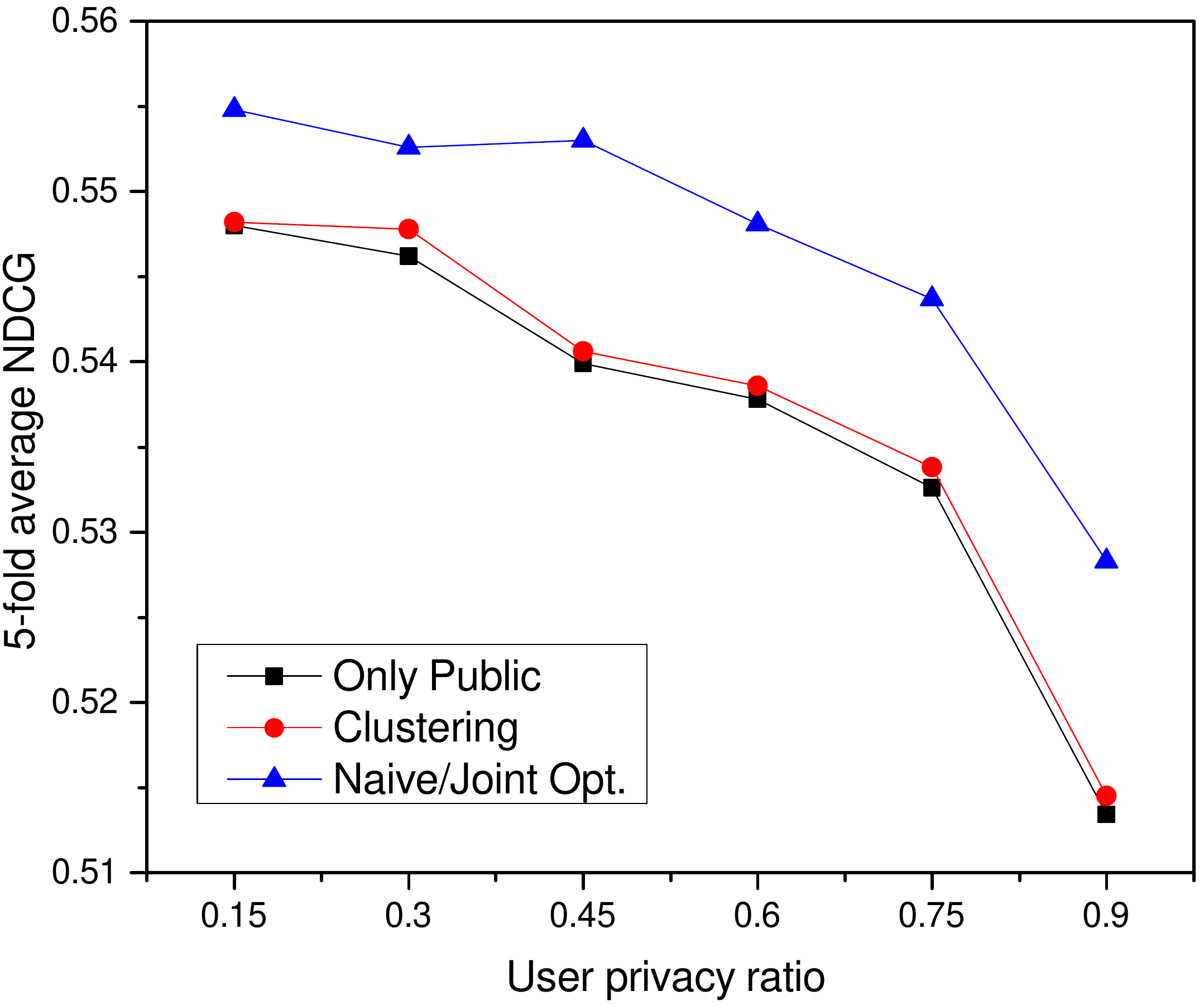}
        \caption{NDCG comparison}
        \label{fig:up_ndcg}
    \end{subfigure}
    \caption{Performance comparison among various baselines for different user privacy ratios on MovieLens dataset.}\label{fig:up_comp}
\end{figure}
\begin{figure}
   \begin{subfigure}[b]{0.48\linewidth}
        \includegraphics[width=\linewidth]{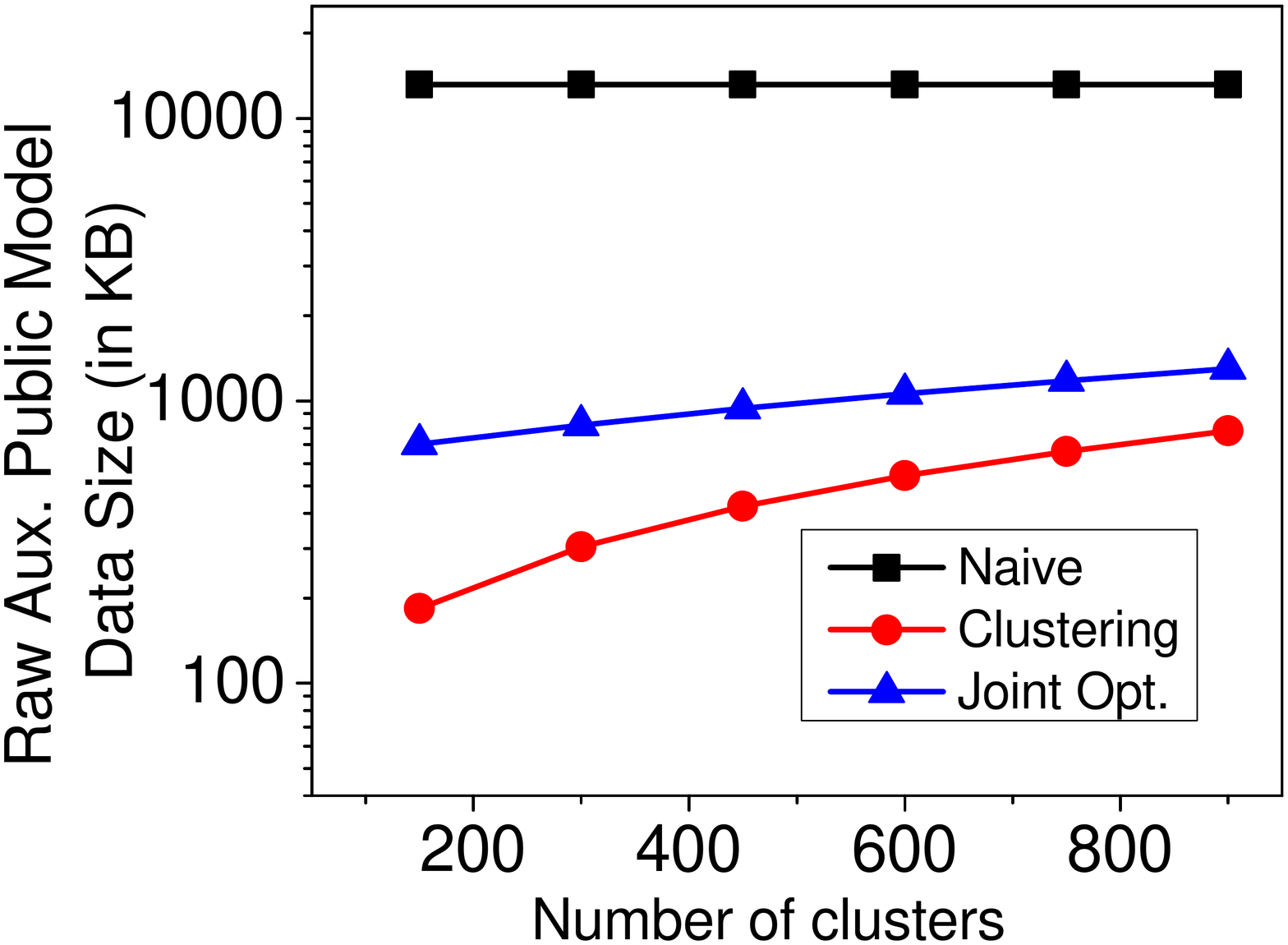}
        \caption{Aux. data size comparison}
        \label{fig:clus_data}
   \end{subfigure}
    \begin{subfigure}[b]{0.48\linewidth}
        \includegraphics[width=\linewidth]{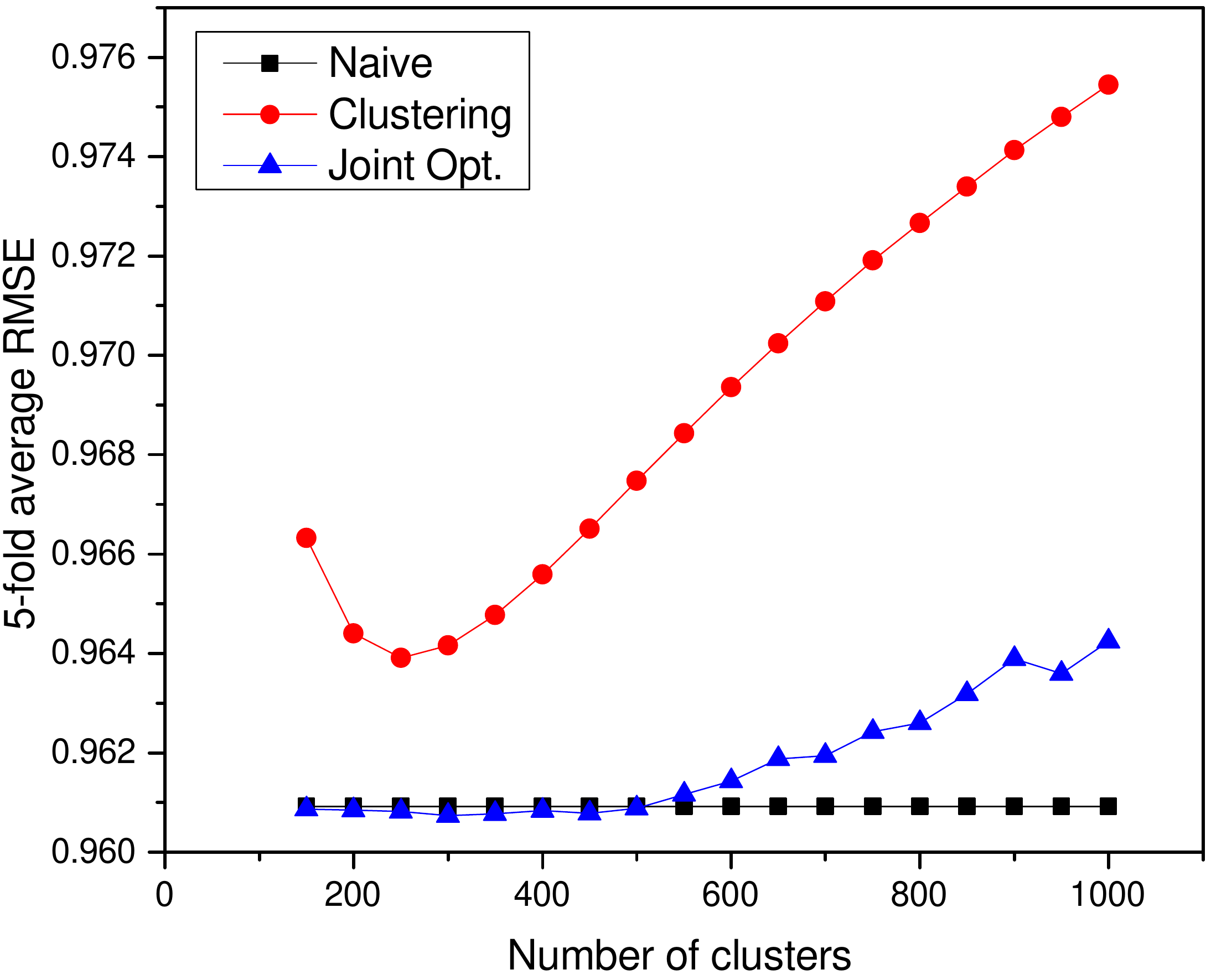}
        \caption{RMSE comparison}
        \label{fig:clus_rmse}
    \end{subfigure}
    \caption{Comparing SP2 implementations for different number of clusters on Amazon Electronics dataset (preprocessed using \emph{mostly balanced} beta distribution and \emph{H1}).}\label{fig:clus_comp}
\end{figure}

As indicated by the results in Table \ref{tab:results}, the peer-to-peer based techniques and the differential privacy method, which attempt to ensure \emph{complete} user privacy from the central recommender system, end up performing worse than the standard \emph{only public} baseline due to the data obfuscation policies. In addition, the fully decentralized approach in \cite{BerkovskyRECSYS07} is not scalable due to the limited number of trusted peers. In the same vein, the distributed aggregation approaches in \cite{ShokriRECSYS09} suffer from poor performance as the number of peers increases due to higher obfuscation; however, lowering the number of peers risks significant privacy breach by the central recommender system.
Table \ref{tab:results} further summarizes that our joint optimization approach (with only top-3 cluster weights) performs as good as the naive approach.
Our clustering approach for SP2 framework, performs worse than naive and joint optimization but is largely better than the \emph{only public} baseline across both evaluation metrics.
Unless otherwise mentioned in the table, $P$-value for all results related to SP2 framework (computed using two-tailed test with respect to \emph{only public} baseline) is less than $0.001$. As evident from the table, our results hold across both the hypotheses. However, the performance of all the implementations improve as the privacy ratio reduces.
This is further demonstrated through figures \ref{fig:up_rmse} and \ref{fig:up_ndcg} which plot the RMSE and NDCG values respectively against varying average user privacy ratio across all users. 
%
Finally, figures \ref{fig:clus_data} and \ref{fig:clus_rmse} present an ablation study that studies the performance and communication cost for different SP2 frameworks with varying number of clusters. The naive method has the best performance but requires the largest auxiliary model data. The joint optimization technique require an order of magnitude less data than the naive one but can reach the same performance for an optimal number of clusters. 
%

\section{Survey}\label{survey}

We conducted a survey\footnote{\url{https://goo.gl/yK2FDd}} to gauge public interest in using our SP2 framework. 
In total, 74 users responded, of which 74\% were male and 24\% were female. 92\% of our respondents were within the age bracket $(18-30)$. 
In our survey, we found that 57\% of the participants do not rate items on any platform, whereas around 20\% of the users provide a lot of ratings. About 48\% of the respondents claim they hesitate to rate an item because they do not want to share their opinion publicly or because they do not trust the platform. 
The last two questions in our survey were aimed at estimating how likely a user is to provide a rating, if he/she can use our selective privacy preserving framework. When users were asked if they would rate more items \emph{privately} on their device, if it guarantees to improve the quality of their recommendations, about 56\% of the users responded affirmatively, while 22\% said `maybe' and 22\% responded with a disagreement. The responses to this survey indicate that an overwhelming majority of users are willing to use our proposed selective privacy preserving framework in order to improve their recommendations as well as safeguard their \emph{private} information. 

\section{Related Work}\label{related}

Privacy preserving recommender systems has been well explored in the literature. 
Peer-to-peer (P2P) techniques \cite{BerkovskyRECSYS07} are largely meant to  
protect users from untrusted servers. However, they also require users to share their private information with peers, which is a privacy breach in itself. 
In addition, P2P architectures lack scalability due to limited number of trusted peers and are vulnerable to malicious interferences by rogue actors. 
Differential privacy methods \cite{MicrosoftDiffPrivacy} provide theoretical privacy guarantees for all users, but can also adversely impact the performance of the recommender systems due to data obfuscation. 

The related literature also comprises of cryptology \cite{zhan2008towards} based techniques that approach the problem little differently. For example, Zhan et al.\cite{zhan2008towards} used ``homomorphic encryption'' to integrate multiple sources of encrypted user ratings in a privacy preserving manner. However, the extreme computation time and scalability issues associated with  homomorphic encryption pose a serious practicality question \cite{HomoEncryPract}, even for moderate size datasets. 
%
%
%
%
%
%

Lastly, recent federated machine learning approaches \cite{OnDevice} have proposed privacy-preserving techniques to build machine learning models using secure aggregation protocol \cite{SecuredAggProtocol}. However, in case of CF algorithms, this would require a user to share an update (in encrypted form) performed on an item factor locally. In our case, this means that the server would be able to identify from the encrypted updates, which items the user had rated privately, even though the exact ratings remain unknown. This itself constitutes a serious privacy breach \cite{ChowICDMW12,Annecdote,VideoPrivacyAct}. Hence, in our SP2 framework, no private user information is ever uploaded or communicated.

\section{Conclusion}\label{conclusion}
In this paper, we proposed a novel selective privacy preserving (SP2) paradigm for CF based recommender systems that allows users to keep a portion of their ratings \emph{private}, meanwhile delivering better recommendations, as compared to other privacy preserving techniques.
We have demonstrated the efficacy of our approach under different configurations by comparing it against other baselines on two real datasets.
Finally, our framework empowers users to define their own privacy policy by determining which ratings should be \emph{private} and which ones should be \emph{public}.

\balance

\bibliographystyle{ACM-Reference-Format}
\bibliography{references}

\end{document}